\documentclass[prl]{revtex4}
\usepackage{graphicx}
\usepackage{bm} 
\usepackage{amssymb} 
\usepackage{amsmath} 

\begin{document}
\relpenalty=9999
\binoppenalty=9999

\title{Erratum: Proposed Chiral Texture of the Magnetic Moments of Unit-Cell Loop Currents in the Pseudogap Phase of Cuprate Superconductors \newline [Phys. Rev. Lett. 111, 047005 (2013)]}

\author{Sergey S. Pershoguba}
\author{Kostyantyn Kechedzhi}
\author{Victor M. Yakovenko}

\affiliation{Condensed Matter Theory Center, Department of Physics, University of Maryland, College Park, Maryland 20742-4111, USA}

\date{\today}

\maketitle

Originally, the observation of the polar Kerr effect in cuprates \cite{Kerr-measurement} was interpreted as the evidence for spontaneous time-reversal-symmetry breaking.  Then, it was proposed in Refs.\ \cite{Hosur-2013,Orenstein-2013,us,Mineev-2013,Karapetyan-2014}, as well as in an earlier paper \cite{Gorkov-1992}, that the polar Kerr effect in cuprates can be explained by a chiral gyrotropic order that breaks inversion symmetry, but preserves time-reversal symmetry.  However, it was shown in a general form using reciprocity arguments by Halperin \cite{Halperin} and confirmed by recent papers \cite{Armitage-2013,Fried-2014} that the reflection matrix of light must be symmetric for a time-reversal-invariant system, so the polar Kerr effect must vanish.  This prompted retractions \cite{Mineev-2014,Hosur-2014} of the proposals that a chiral order without time-reversal-symmetry breaking can explain the polar Kerr effect.  In this Erratum, we show that, while the electromagnetic constituent relations
are correctly derived in our Letter \cite{us} and do contain a bulk gyrotropic term, the reflection matrix of light is, nevertheless, symmetric (in agreement with Refs.\ \cite{Halperin,Armitage-2013,Fried-2014,Mineev-2014,Hosur-2014}), so our proposed model \cite{us} cannot explain the experimental observation \cite{Kerr-measurement} of the polar Kerr effect in cuprates \cite{Faraday}.

The confusion stems from different treatments of a surface contribution to the constituent relations in different papers.  Reference \cite{Gorkov-1992} employed the bulk relation $4\pi\bm P=\gamma\bm\nabla\times\bm E$, where $\bm E$ and $\bm P$ are the electric field and polarization, and $\gamma$ is the coefficient of natural optical activity \cite{Landau}.  However, for a system occupying semi-infinite space $z>0$ in contact with vacuum at $z<0$, the coefficient $\gamma(z)$ has a stepwise dependence on coordinate $z$: $\gamma(z)=0$ for $z<0$ and $\gamma(z)\neq0$ for $z>0$. Reference~\cite{Zheludev} proposed the following relation $4\pi\bm P=\bm\nabla\times(\gamma\bm E)=\gamma\bm\nabla\times\bm E+(\bm\nabla_z\gamma)\times\bm E$ containing the delta-function surface term $\gamma\delta(z)\hat{\bm z}\times\bm E$.  Substituting these relations into Maxwell's equations, Refs.\ \cite{Gorkov-1992} and \cite{Zheludev} obtained opposite signs for the polar Kerr effect.  However, both relations are wrong, as pointed out in Ref.\ \cite{Fried-2014}, and the correct relation is $4\pi\bm P=\gamma\bm\nabla\times\bm E+(1/2)(\bm\nabla_z\gamma)\times\bm E$, as employed in Ref.\ \cite{Mineev-2014} after correcting an arithmetic error in Ref.\ \cite{Mineev-2010}.  This relation can be obtained by variation $\bm P=\delta S/\delta\bm E$ of the effective action $S=(1/8\pi)\int d\omega\,d^3r\,\gamma(z)\,\bm E \cdot(\bm\nabla\times \bm E)$, and it gives zero polar Kerr effect \cite{Mineev-2014}.

Equation (4) in our Letter \cite{us} utilized the incorrect formula from Ref.~\cite{Zheludev} claiming a nonzero Kerr angle.  However, our microscopic derivation of the effective action for a helical structure of loop currents is correct.  Moreover, the advantage of our discrete lattice model over continuous models is that the correct surface term in the constituent relations can be derived unambiguously without confusion.  The electromagnetic action in our model is given by Eq.~(9) in our Letter~\cite{us}
\begin{equation}
	S =-\frac{\tilde\gamma}{4\pi}\sum_{n=0}^\infty\int d\omega\,d^2r\,(\bm E_n\times\bm N_n)\cdot(\bm E_{n+1}\times\bm N_{n+1}), 
	\label{action}
\end{equation}
where $n$ is an integer coordinate labeling cuprate layers in the $z$ direction, and the parameter $\tilde\gamma=4\pi\Lambda\beta^2$ encapsulates the magnetoelectric coefficient $\beta$ and the magnetic coupling $\Lambda$.  The in-plane anapole vectors $\bm N_n$ are arranged in a helical structure, so that $\bm N_{n+1}$ is rotated by $\pi/2$ around the $z$ axis relative to $\bm N_{n}$ (i.e. $\bm N_0 = \hat{\bm x} N$, $\bm N_1 = \hat{\bm y} N$, $\bm N_2 = -\hat{\bm  x}N,\,\ldots$). Although the vectors $\bm N_n$ change sign upon the time-reversal operation, the action~(\ref{action}) is time-reversal invariant, because it is bilinear in $\bm N_n$.  
The action (\ref{action}) generates different expressions for the electric polarizations $\bm P_n$ in the bulk for $n>0$ and $\bm P_0$ at the surface layer at $n=0$, because the latter has only one neighboring layer:
\begin{equation}
  \bm P_n = \frac1d \frac{\delta S}{\delta\bm E_n} = \frac{\tilde\gamma}{4\pi d}\,
  \bm N_{n+1}\left[\bm N_n\cdot\left(\bm E_{n+1}-\bm E_{n-1}\right) \right], 
  \qquad \bm P_0 = \frac1d \frac{\delta S}{\delta\bm E_0} 
  =\frac{\tilde\gamma}{4\pi d}\,\bm N_{1}\left(\bm N_0\cdot\bm E_{1}\right),
\label{pol}
\end{equation}
where $d$ is the interlayer distance, and we used the property $\bm N_{n+2}=-\bm N_n$.  In the continuum limit $d\to0$, the first and the second terms in Eq.~(\ref{pol}) produce the bulk and the surface contributions to the electric polarization, respectively,
\begin{equation}
  4\pi\bm P = \gamma\,\bm\nabla_z\times\bm E 
  + \frac12 \gamma\,\delta(z)\,(\tau_x-i\tau_y) \bm E,
\label{polCont}
\end{equation}
where $\gamma=\tilde\gamma\,\hat{\bm z}\cdot[\bm N^{(n)}\times\bm N^{(n+1)}]$, and the Pauli matrices $\tau_x$ and $\tau_y$ act on the two-component electric field $\bm E=(E_x,E_y)$.
In agreement with the discussion above, the surface term in Eq.~(\ref{polCont}) contains the antisymmetric contribution proportional to $i\tau_y$ with the coefficient 1/2 relative to the bulk term, as in Eq.~(6) of Ref.~\cite{Mineev-2014}.  The surface term in Eq.~(\ref{polCont}) also contains the symmetric contribution proportional to $\tau_x$, which represents nematicity in our lattice model.  Substituting Eq.~(\ref{polCont}) into Maxwell's equations with the electric current density given by $\bm j=\dot{\bm P}$, we find that the plane-wave eigenmodes propagating in the $z$ direction with the frequency $\omega$ are circularly polarized $\bm E_\pm\propto(1,\pm i)$ with the momenta $k_\pm=k(1\pm k\gamma/2)$, where $k=\omega/c$.  By matching the incoming $\bm E_Ie^{ikz} $ and reflected $\bm E_Re^{-ikz}$  waves with the eigenmodes $\bm E_\pm$ in the bulk and using the correct boundary condition determined by the surface term in Eq.~(\ref{polCont}), we find, indeed, that the reflection matrix is symmetric,  so the polar Kerr effect vanishes.

The same conclusion can be also obtained without taking the continuous limit.   The electric polarization (\ref{pol}) gives the following contribution to the right-hand side of Maxwell's equation for the waves propagating in the $z$ direction
\begin{equation}
  \nabla_z^2\bm E + k^2\bm E = -4\pi k^2d\sum_{n=0}^\infty\bm P_n\,\delta(z-nd). 
\label{maxwell}
\end{equation}
Treating the right-hand side of Eq.~(\ref{maxwell}) as a perturbation \cite{gf} up to the first order in $\gamma$, we obtain the reflection matrix
\begin{equation}
	\bm E_R =  \frac{i\gamma k}{4\cos kd}\,\tau_x\,\bm E_I,
	\label{reflected}
\end{equation}
which contains only the symmetric matrix $\tau_x$, so the polar Kerr effect vanishes.

In conclusion, we withdraw our claim made in our Letter~\cite{us} that the chiral texture of loop currents can explain the experimentally observed polar Kerr effect in cuprates \cite{Kerr-measurement}.  Although our model does contain a bulk gyrotropic term and produces a nonzero Faraday effect on transmission, the reflection matrix of light is, nevertheless, symmetric and gives zero polar Kerr effect.  The correct result is obtained when the surface term is properly derived from the discrete lattice model in out Letter \cite{us}.  Incorrect conclusions about a nonzero polar Kerr effect were made when the surface term was either omitted \cite{Gorkov-1992} or had an incorrect factor \cite{Zheludev}.  A theoretical explanation of the experimental results \cite{Kerr-measurement} still remains an open question. An alternative approach using a time-reversal-breaking tilted loop-current model was proposed in Ref.~\cite{Yakovenko-2014}.

This work was supported by JQI-NSF-PFC (K.~K.).


\begin{thebibliography}{9}
\bibitem{Faraday} While the polar Kerr effect for reflection of light vanishes for a chiral time-reversal-invariant system, the Faraday effect for transmission of light does not vanish, as it is commonly accepted, e.g.,\ in Ref.\ \cite{Landau}.
\bibitem{gf} We use Green's function $G(z)=e^{ik|z|}/2ik$, which satisfies the equation 
$\nabla_z^2 G(z) + k^2 G(z) = \delta(z)$.  In the spirit of the first Born approximation, we substitute the zeroth-order electric field $\bm E_n = \bm E_Ie^{iknd}$ into the right-hand side of Eq.~(\ref{maxwell}) and compute the first-order correction in $\gamma$: $\bm E^{(1)}(z) = -4\pi dk^2\int dz' \,G(z-z') \sum\limits_{n=0}^\infty\bm P_n\,\delta(z'-nd)$. Then, the reflected wave is $\bm E_R = E^{(1)}(0)$.
\end{thebibliography}
\end{document}